%
%
%
%
%
\input harvmac
\edef\cite{\the\catcode`@}%
\catcode`@ = 11
\let\@oldatcatcode = \cite
\chardef\@letter = 11
\chardef\@other = 12
%
%
%
%
\def\@innerdef#1#2{\edef#1{\expandafter\noexpand\csname #2\endcsname}}%
%
%
\@innerdef\@innernewcount{newcount}%
\@innerdef\@innernewdimen{newdimen}%
\@innerdef\@innernewif{newif}%
\@innerdef\@innernewwrite{newwrite}%
%
%
%
\def\@gobble#1{}%
%
%
%
\ifx\inputlineno\@undefined
   \let\@linenumber = \empty 
\else
   \def\@linenumber{\the\inputlineno:\space}%
\fi
%
%
%
\def\@futurenonspacelet#1{\def\cs{#1}%
   \afterassignment\@stepone\let\@nexttoken=
}%
\begingroup 
\def\\{\global\let\@stoken= }%
\\ 
\endgroup
\def\@stepone{\expandafter\futurelet\cs\@steptwo}%
\def\@steptwo{\expandafter\ifx\cs\@stoken\let\@@next=\@stepthree
   \else\let\@@next=\@nexttoken\fi \@@next}%
\def\@stepthree{\afterassignment\@stepone\let\@@next= }%
%
%
%
\def\@getoptionalarg#1{%
   \let\@optionaltemp = #1%
   \let\@optionalnext = \relax
   \@futurenonspacelet\@optionalnext\@bracketcheck
}%
%
%
\def\@bracketcheck{%
   \ifx [\@optionalnext
      \expandafter\@@getoptionalarg
   \else
      \let\@optionalarg = \empty
      \expandafter\@optionaltemp
   \fi
}%
\def\@@getoptionalarg[#1]{%
   \def\@optionalarg{#1}%
   \@optionaltemp
}%
%
%
%
\def\@nnil{\@nil}%
\def\@fornoop#1\@@#2#3{}%
\def\@for#1:=#2\do#3{%
   \edef\@fortmp{#2}%
   \ifx\@fortmp\empty \else
      \expandafter\@forloop#2,\@nil,\@nil\@@#1{#3}%
   \fi
}%
\def\@forloop#1,#2,#3\@@#4#5{\def#4{#1}\ifx #4\@nnil \else
       #5\def#4{#2}\ifx #4\@nnil \else#5\@iforloop #3\@@#4{#5}\fi\fi
}%
\def\@iforloop#1,#2\@@#3#4{\def#3{#1}\ifx #3\@nnil
       \let\@nextwhile=\@fornoop \else
      #4\relax\let\@nextwhile=\@iforloop\fi\@nextwhile#2\@@#3{#4}%
}%
%
%
%
\@innernewif\if@fileexists
\def\@testfileexistence{\@getoptionalarg\@finishtestfileexistence}%
\def\@finishtestfileexistence#1{%
   \begingroup
      \def\extension{#1}%
      \immediate\openin0 =
         \ifx\@optionalarg\empty\jobname\else\@optionalarg\fi
         \ifx\extension\empty \else .#1\fi
         \space
      \ifeof 0
         \global\@fileexistsfalse
      \else
         \global\@fileexiststrue
      \fi
      \immediate\closein0
   \endgroup
}%
%
%
%
%
\def\bibliographystyle#1{%
   \@readauxfile
   \@writeaux{\string\bibstyle{#1}}%
}%
\let\bibstyle = \@gobble
%
%
\let\bblfilebasename = \jobname
\def\bibliography#1{%
   \@readauxfile
   \@writeaux{\string\bibdata{#1}}%
   \@testfileexistence[\bblfilebasename]{bbl}%
   \if@fileexists
      \nobreak
      \@readbblfile
   \fi
}%
\let\bibdata = \@gobble
%
%
\def\nocite#1{%
   \@readauxfile
   \@writeaux{\string\citation{#1}}%
}%
\@innernewif\if@notfirstcitation
%
%
\def\cite{\@getoptionalarg\@cite}%
%
%
\def\@cite#1{%
   \let\@citenotetext = \@optionalarg
   \printcitestart
   \nocite{#1}%
   \@notfirstcitationfalse
   \@for \@citation :=#1\do
   {%
      \expandafter\@onecitation\@citation\@@
   }%
   \ifx\empty\@citenotetext\else
      \printcitenote{\@citenotetext}%
   \fi
   \printcitefinish
}%
\def\@onecitation#1\@@{%
   \if@notfirstcitation
      \printbetweencitations
   \fi
   \expandafter \ifx \csname\@citelabel{#1}\endcsname \relax
      \if@citewarning
         \message{\@linenumber Undefined citation `#1'.}%
      \fi
      \expandafter\gdef\csname\@citelabel{#1}\endcsname{%
         {\tt
            \escapechar = -1
            \nobreak\hskip0pt
            \expandafter\string\csname#1\endcsname
            \nobreak\hskip0pt
         }%
      }%
   \fi
   \@printcitelabel{#1}%
   \@notfirstcitationtrue
}%
%
\def\@printcitelabel#1{%
   \csname\@citelabel{#1}\endcsname
}%
%
%
\def\@citelabel#1{b@#1}%
%
%
\def\@citedef#1#2{\expandafter\gdef\csname\@citelabel{#1}\endcsname{#2}}%
%
%
%
\def\@readbblfile{%
   \ifx\@itemnum\@undefined
      \@innernewcount\@itemnum
   \fi
   \begingroup
      \ifx\begin\undefined
         \def\begin##1##2{%
            \setbox0 = \hbox{\biblabelcontents{##2}}%
            \biblabelwidth = \wd0
         }%
         \let\end = \@gobble 
      \fi
      %
      %
      \@itemnum = 0
      \def\bibitem{\@getoptionalarg\@bibitem}%
      \def\@bibitem{%
         \ifx\@optionalarg\empty
            \expandafter\@numberedbibitem
         \else
            \expandafter\@alphabibitem
         \fi
      }%
      \def\@alphabibitem##1{%
         \expandafter \xdef\csname\@citelabel{##1}\endcsname {\@optionalarg}%
         \ifx\biblabelprecontents\@undefined
            \let\biblabelprecontents = \relax
         \fi
         \ifx\biblabelpostcontents\@undefined
            \let\biblabelpostcontents = \hss
         \fi
         \@finishbibitem{##1}%
      }%
      \def\@numberedbibitem##1{%
         \advance\@itemnum by 1
         \expandafter \xdef\csname\@citelabel{##1}\endcsname{\number\@itemnum}%
         \ifx\biblabelprecontents\@undefined
            \let\biblabelprecontents = \hss
         \fi
         \ifx\biblabelpostcontents\@undefined
            \let\biblabelpostcontents = \relax
         \fi
         \@finishbibitem{##1}%
      }%
      \def\@finishbibitem##1{%
         \biblabelprint{\csname\@citelabel{##1}\endcsname}%
         \@writeaux{\string\@citedef{##1}{\csname\@citelabel{##1}\endcsname}}%
         \ignorespaces
      }%
      %
      %
      \let\em = \bblem
      \let\newblock = \bblnewblock
      \let\sc = \bblsc
      \frenchspacing
      \clubpenalty = 4000 \widowpenalty = 4000
      \tolerance = 10000 \hfuzz = .5pt
      \everypar = {\hangindent = \biblabelwidth
                      \advance\hangindent by \biblabelextraspace}%
      \bblrm
      \parskip = 1.5ex plus .5ex minus .5ex
      \biblabelextraspace = .5em
      \bblhook
      \input \bblfilebasename.bbl
   \endgroup
}%
%
%
\@innernewdimen\biblabelwidth
\@innernewdimen\biblabelextraspace
%
%
%
\def\biblabelprint#1{%
   \noindent
   \hbox to \biblabelwidth{%
      \biblabelprecontents
      \biblabelcontents{#1}%
      \biblabelpostcontents
   }%
   \kern\biblabelextraspace
}%
%
%
%
\def\biblabelcontents#1{{\bblrm [#1]}}%
%
%
\def\bblrm{\rm}%
%
%
\def\bblem{\it}%
%
%
\def\bblsc{\ifx\@scfont\@undefined
              \font\@scfont = cmcsc10
           \fi
           \@scfont
}%
%
%
\def\bblnewblock{\hskip .11em plus .33em minus .07em }%
%
%
\let\bblhook = \empty
%
%
%
\def\printcitestart{[}
\def\printcitefinish{]}
\def\printbetweencitations{, }
\def\printcitenote#1{, #1}
%
%
%
\let\citation = \@gobble
%
%
%
\@innernewcount\@numparams
%
%
\def\newcommand#1{%
   \def\@commandname{#1}%
   \@getoptionalarg\@continuenewcommand
}%
%
%
\def\@continuenewcommand{%
   \@numparams = \ifx\@optionalarg\empty 0\else\@optionalarg \fi \relax
   \@newcommand
}%
%
%
\def\@newcommand#1{%
   \def\@startdef{\expandafter\edef\@commandname}%
   \ifnum\@numparams=0
      \let\@paramdef = \empty
   \else
      \ifnum\@numparams>9
         \errmessage{\the\@numparams\space is too many parameters}%
      \else
         \ifnum\@numparams<0
            \errmessage{\the\@numparams\space is too few parameters}%
         \else
            \edef\@paramdef{%
               \ifcase\@numparams
                  \empty  No arguments.
               \or ####1%
               \or ####1####2%
               \or ####1####2####3%
               \or ####1####2####3####4%
               \or ####1####2####3####4####5%
               \or ####1####2####3####4####5####6%
               \or ####1####2####3####4####5####6####7%
               \or ####1####2####3####4####5####6####7####8%
               \or ####1####2####3####4####5####6####7####8####9%
               \fi
            }%
         \fi
      \fi
   \fi
   \expandafter\@startdef\@paramdef{#1}%
}%
%
%
%
%
\def\@readauxfile{%
   \if@auxfiledone \else 
      \global\@auxfiledonetrue
      \@testfileexistence{aux}%
      \if@fileexists
         \begingroup
            \endlinechar = -1
            \catcode`@ = 11
            \input \jobname.aux
         \endgroup
      \else
         \message{\@undefinedmessage}%
         \global\@citewarningfalse
      \fi
      \immediate\openout\@auxfile = \jobname.aux
   \fi
}%
%
%
\newif\if@auxfiledone
\ifx\noauxfile\@undefined \else \@auxfiledonetrue\fi
%
%
%
%
\@innernewwrite\@auxfile
\def\@writeaux#1{\ifx\noauxfile\@undefined \write\@auxfile{#1}\fi}%
%
%
%
\ifx\@undefinedmessage\@undefined
   \def\@undefinedmessage{No .aux file; I won't give you warnings about
                          undefined citations.}%
\fi
%
%
\@innernewif\if@citewarning
\ifx\noauxfile\@undefined \@citewarningtrue\fi
%
%
%
\catcode`@ = \@oldatcatcode
\def\href#1#2{#2}  
\def\Introduction{1}
\def\fields{(1)}
\def\ws{(2)}
\def\gaugeop{(3)}
\def\stringop{(4)}
\def\dilvert{(5)}
\def\gravvert{(6)}
\def\confan{(7)}
\def\DThreeGeom{(8)}
\def\DThreeVars{(9)}
\def\NearGeom{(10)}
\def\zphi{(11)}
\def\Ident{(12)}
\def\BndCond{(13)}
\def\SymBC{1}
\def\AnyADS{(14)}
\def\AllowedSize{(15)}
\def\AllowedZeta{(16)}
\def\LieVar{(17)}
\def\CKE{(18)}
\def\TwoPt{2}
\def\MinAction{(19)}
\def\MinEOM{(20)}
\def\CompleteSet{(21)}
\def\WVCoup{(22)}
\def\SimpIdent{(23)}
\def\KBound{(24)}
\def\FluxDef{(25)}
\def\TwoPtO{(26)}
\def\EvalFlux{(27)}
\def\poscorr{(28)}
\def\TwoPtFct{(29)}
\def\XFourDef{(30)}
\def\SInt{(31)}
\def\TwoTT{(32)}
\def\MassiveString{3}
\def\PartAction{(33)}
\def\PartEOM{(34)}
\def\PartSols{(35)}
\def\PhiBehave{(36)}
\def\GotFlux{(37)}
\def\higher{(38)}
\def\Coulone{(39)}
\def\Coultwo{(40)}
\def\anom{(41)}
\def\massivewave{(42)}
\def\genexpand{(43)}
\def\OOFct{(44)}
\def\opdim{(45)}
\def\anomform{(46)}

\writedefs
\sequentialequations


\def\rf#1{\csname #1\endcsname{}}

\def\TL{\hfil$\displaystyle{##}$}
\def\TR{$\displaystyle{{}##}$\hfil}




\def\comment#1{}
\def\fixit#1{}

\def\tf#1#2{{\textstyle{#1 \over #2}}}


\def\tr{\mathop{\rm tr}\nolimits}



\def\lsim{\mathrel{\mathstrut\smash{\ooalign{\raise2.5pt\hbox{$<$}\cr\lower2.5pt\hbox{$\sim$}}}}}
\def\gsim{\mathrel{\mathstrut\smash{\ooalign{\raise2.5pt\hbox{$>$}\cr\lower2.5pt\hbox{$\sim$}}}}}



\def\sqr#1#2{{\vcenter{\vbox{\hrule height.#2pt
         \hbox{\vrule width.#2pt height#1pt \kern#1pt
            \vrule width.#2pt}
         \hrule height.#2pt}}}}
\def\square{\mathop{\mathchoice\sqr56\sqr56\sqr{3.75}4\sqr34}\nolimits}





\def\+{^\dagger}




\def\overleftrightarrow#1{\vbox{\ialign{##\crcr
     $\leftrightarrow$\crcr\noalign{\kern-0pt\nointerlineskip}
     $\hfil\displaystyle{#1}\hfil$\crcr}}}

\def\em{\it}   


\Title{
 \vbox{\baselineskip10pt
  \hbox{PUPT-1767}
  \hbox{hep-th/9802109}
 }
}
{
 \vbox{
  \centerline{Gauge Theory Correlators from Non-Critical String Theory}
 }
}
\vskip -25 true pt

\centerline{
 S.S.~Gubser,\footnote{$^1$}{e-mail:  ssgubser@viper.princeton.edu}
 I.R.~Klebanov\footnote{$^2$}{e-mail:  klebanov@viper.princeton.edu}
and
 A.M.~Polyakov\footnote{$^3$}{e-mail:  polyakov@puhep1.princeton.edu}
}

\centerline{\it Joseph Henry Laboratories, 
Princeton University, Princeton, NJ  08544}

\bigskip\bigskip

\centerline {\bf Abstract}
\smallskip
\baselineskip12pt
\noindent

We suggest a means of obtaining certain Green's functions in
$3+1$-dimensional ${\cal N} = 4$ supersymmetric 
Yang-Mills theory with a large number of colors via non-critical
string theory. The non-critical string theory
is related to critical string theory in anti-deSitter background.
We introduce a boundary of the
anti-deSitter space analogous to a cut-off on
the Liouville coordinate of the two-dimensional
string theory. 
Correlation functions of operators in the gauge
theory are related to the dependence of the
supergravity action on the boundary conditions.
From the quadratic terms in supergravity 
we read off the anomalous dimensions.
For operators that couple to massless string states it has been
established through absorption calculations that the anomalous
dimensions vanish, and we rederive this result. 
The operators that couple to massive string states at level $n$
acquire anomalous dimensions that grow as
$2\left (n g_{YM} \sqrt {2 N} \right )^{1/2}$  for large `t Hooft coupling.
This is a new prediction about the strong coupling behavior
of large $N$ SYM theory.

\Date{February 1998}

\noblackbox
\baselineskip 14pt plus 1pt minus 1pt


\newsec{Introduction}
\seclab\Introduction

Relations between gauge fields and strings present an old, fascinating
and unanswered question. The full answer to this question is of great
importance for theoretical physics. It will provide us with a
theory of quark confinement  by explaining the dynamics of
color-electric fluxes. On the other hand, it will perhaps uncover
the true ``gauge'' degrees of freedom of the fundamental string theories,
and therefore of gravity.

The Wilson loops of gauge theories satisfy the loop equations
which translate the Schwinger-Dyson equations into variational
equations on the loop space \cite{AP,mm}. 
These equations should have a solution
in the form of the sum over random surfaces bounded by the loop.
These are the world surfaces of the color-electric fluxes.
For the $SU(N)$ Yang-Mills theory they are expected to carry the 
`t Hooft factor \cite{GT}, $N^\chi$, where $\chi$ is the Euler
character. Hence, in the large $N$ limit 
where $g_{YM}^2 N$ is kept fixed only the simplest
topologies are relevant.

Until recently, the action for the ``confining string''
had not been known. In \cite{Sasha} is was suggested that it
must have a rather unusual structure. Let us describe it briefly.
First of all, the world surface of the electric flux propagates
in at least 5 dimensions. This is because the non-critical strings
are described by the fields
\eqn\fields{X^\mu (\sigma)\ ,\qquad g_{ij} (\sigma) =
e^{\varphi (\sigma)} \delta_{ij}\ ,
}
where $X^\mu$ belong to 4-dimensional (Euclidean) space and
$g_{ij} (\sigma)$ is the world sheet metric in the conformal gauge.
The general form of the world sheet lagrangian compatible with
the 4-dimensional symmetries is
\eqn\ws{
{\cal L}= {1\over 2} (\partial_i \varphi)^2 + a^2 (\varphi)
(\partial_i X^\mu)^2 + \Phi (\varphi) ^{(2)}R + {\rm Ramond-Ramond
\ backgrounds}\ ,
}
where $^{(2)}R$ is the world sheet curvature, $\Phi(\varphi)$
is the dilaton \cite{ft}, while the field 
$$ \Sigma (\varphi) = a^2 (\varphi)
$$
defines a variable string tension. In order to reproduce the zig-zag
symmetry of the Wilson loop, the gauge fields must be
located at a certain value $\varphi=\varphi_*$ such 
that $a(\varphi_*)=0$.
We will call this point ``the horizon.''

The background fields $\Phi(\varphi)$, $a(\varphi)$ and others must be
chosen to satisfy the conditions of conformal invariance on the
world sheet \cite{cmpf}. 
After this is done, the relation between gauge fields and 
strings can be described as an isomorphism between the general Yang-Mills
operators of the type
\eqn\gaugeop{
\int d^4 x e^{ip\cdot x}\tr \left (\nabla_{\alpha_1}\ldots F_{\mu_1 \nu_1}
\ldots \nabla_{\alpha_n}\ldots F_{\mu_n \nu_n} (x)\right )
}
and the algebra of vertex operators of string theory, which have the form
\eqn\stringop{ V^{\alpha_1 \ldots \alpha_n} (p)= \int d^2 \sigma 
\Psi_p^{i_1\ldots i_n j_1\ldots j_m} \big (\varphi (\sigma) \big )
e^{ip\cdot X(\sigma)} \partial_{i_1} X^{\alpha_1}
\ldots \partial_{i_n} X^{\alpha_n}\partial_{j_1}\varphi\ldots
 \partial_{j_m} \varphi  
\ ,
}
where the wave functions $\Psi_p^{i_1\ldots i_n j_1\ldots j_m} (\varphi)$
are again determined by the conformal invariance on the world sheet.
The isomorphism mentioned above implies the coincidence of the correlation 
functions of these two sets of vertex operators.

Another, seemingly unrelated, development is connected with the
Dirichlet brane \cite{brane} description of black 3-branes
in \cite{gkp,kleb,gukt,gkThree}. 
The essential observation is that, on the one hand, the black
branes
are solitons which curve space \cite{hs} 
and, on the other hand, the world volume of $N$
parallel D-branes is described by supersymmetric $U(N)$ gauge theory with
16 supercharges \cite{Witten}.
A particularly interesting system is 
provided by the limit of
a large number $N$ of coincident D3-branes \cite{gkp,kleb,gukt,gkThree}, 
whose world volume is described by ${\cal N}=4$ supersymmetric $U(N)$ 
gauge theory in $3+1$ dimensions. For large $g_{YM}^2 N$ the
curvature of the classical geometry becomes small compared to the
string scale \cite{kleb}, 
which allows for comparison
of certain correlation functions between the supergravity and the gauge
theory, with perfect agreement \cite{kleb,gukt,gkThree}. 
Corrections in powers of $\alpha'$ times the curvature
on the string theory side correspond to corrections in powers of
$(g_{YM}^2 N)^{-1/2}$ on the gauge theory side.
The string loop corrections are suppressed by powers 
of $1/N^2$.

The vertex operators introduced in \cite{kleb,gukt,gkThree} describe
the coupling of massless closed string fields to the world volume.
For example, the vertex operator for the dilaton is
\eqn\dilvert{ 
{1\over 4 g_{YM}^2}
\int d^4 x e^{ip\cdot x}
\left (\tr F_{\mu\nu} F^{\mu\nu} (x)+ \ldots \right )\ , }
while that for the graviton polarized along the 3-branes is
\eqn\gravvert{ \int d^4 x e^{ip\cdot x} T^{\mu\nu}(x)\ ,
}
where $T^{\mu\nu}$ is the stress tensor.
The low energy absorption cross-sections are related to the
2-point functions of the vertex operators, and turn out to be
in complete agreement with conformal invariance and supersymmetric
non-renormalization theorems \cite{gkThree}. An earlier
calculation of the entropy as a function of temperature 
for $N$ coincident D3-branes \cite{gkp} exhibits a dependence 
expected of a field theory with $O(N^2)$ massless fields, 
and turns out to be $3/4$ of the free field answer. 
This is not a discrepancy since the free field result is valid
for small $g_{YM}^2 N$, while the result of \cite{gkp} is applicable as
$g_{YM}^2 N\rightarrow \infty$. We
now regard this result as a non-trivial prediction of
supergravity concerning the strong coupling behavior of ${\cal N} = 4$
supersymmetric gauge theory at large $N$ and finite temperature.

The non-critical string and the D-brane approaches to 3+1
dimensional gauge theory have been synthesized in \cite{jthroat}
by rescaling the 3-brane metric and taking the limit in which it has 
conformal symmetry, being the direct product $AdS_5\times S^5$.
\foot{The papers that put an early
emphasis on the anti-deSitter nature of the near-horizon region of
certain brane configurations, 
and its relation with string and M-theory, are
\cite{gt,ght}. Other ideas on the relation between branes and AdS
supergravity were recently pursued in
\cite{Hyun,Boonstra,Sfetsos}.}
This is exactly the confining string ansatz \cite{Sasha} with
\eqn\confan{ a(\varphi) = e^{\varphi/R}\ ,
}
corresponding to the case of constant negative curvature of order
$1/R^2$. The horizon is located at $\varphi_*=-\infty$.
The Liouville field is thus related to the radial coordinate
of the space transverse to the 3-brane. The extra $S^5$
part of the metric is associated with the 6 scalars and the $SU(4)$
R-symmetry present in the 
${\cal N}=4$ supersymmetric gauge theory.

In the present paper we
make the next step and show how the 
masses of excited states of the ``confining
string'' are related to the anomalous dimensions of the SYM theory.
Hopefully this analysis will help future explorations of asymptotically
free gauge theories needed for quark confinement. 

We will suggest a potentially very rich and detailed means of
analyzing the throat-brane correspondence: we propose an
identification of the generating function of the Green's functions of
the superconformal world-volume theory and the supergravity action in
the near horizon background geometry.
We will find it necessary to introduce a boundary of the $AdS_5$
space near the place where the throat turns into the asymptotically
flat space. Thus, the anti-deSitter coordinate $\varphi$ is 
defined on a half-line $(-\infty, 0]$,
similarly to the Liouville coordinate of the 2-dimensional string
theory \cite{Polch,DJ}. 
The correlation functions are specified by the dependence of
the action on the boundary conditions, 
again in analogy with the $c=1$ case. 
One new prediction that we will be able to extract this way is for the 
anomalous dimensions of the gauge
theory operators that correspond to massive string
states. For a state at level $n$ we find that, for large 
$g_{YM}^2 N$, the anomalous dimension grows as
$2 \sqrt n (2 g_{YM}^2 N)^{1/4}$.

\newsec{Green's functions from the supergravity action}

The geometry of a large number $N$ of coincident D3-branes is
  \eqn\DThreeGeom{
   ds^2 = \left( 1 + {R^4 \over r^4} \right)^{-1/2}
    \left( -dt^2 + d\vec{x}^2 \right) + 
    \left( 1 + {R^4 \over r^4} \right)^{1/2} 
    \left( dr^2 + r^2 d\Omega_5^2 \right) \ .
  }
 The parameter $R$, where
  \eqn\DThreeVars{
   R^4 = {N \over 2\pi^2 T_3}\ , \qquad\qquad 
   T_3 = {\sqrt{\pi} \over \kappa}
  }
 is the only length scale involved in all of what we will say.  $T_3$
is the tension of a single D3-brane, and $\kappa$ is the
ten-dimensional gravitational coupling.  The near-horizon geometry of
$N$ D3-branes is $AdS_5 \times S^5$, as one can see most easily by
defining the radial coordinate $z = R^2/r$.  Then 
  \eqn\NearGeom{
   ds^2 = {R^2 \over z^2} \left( -dt^2 + d\vec{x}^2 + dz^2 \right) + 
    R^2 d\Omega_5^2 \ .
  }
The relation to the coordinate $\varphi$ used in the previous section
is
\eqn\zphi{ z= R e^{-\varphi/R}\ .
}
 Note that the limit $z \to 0$ is far from the brane.  Of course, for
$z \lsim R$ the $AdS$ form \NearGeom\ gets modified, and for $z \ll R$
one obtains flat ten-dimensional Minkowski space.  We will freely use
phrases like ``far from the brane'' and ``near the brane'' to describe
regimes of small $z$ and large $z$, despite the fact that the geometry
is geodesically complete and nonsingular.

The basic idea is to identify the generating functional of connected
Green's functions in the gauge theory with the minimum of the
supergravity action, subject to some boundary conditions at 
$z=R$ and $z=\infty$:
  \eqn\Ident{
   W[g_{\mu\nu}(x^\lambda)] = K[g_{\mu\nu}(x^\lambda)] = 
     S[g_{\mu\nu}(x^\lambda,z)] \ .
  }
 $W$ generates the connected Green's functions of the gauge theory; $S$ is 
the supergravity action on the $AdS$ space; while $K$ is the minimum of $S$ 
subject to the boundary conditions.
 We have kept only the metric $g_{\mu\nu}(x^\lambda)$ of the
world-volume as an explicit argument of $W$.  The boundary conditions
subject to which the supergravity action $S$ is minimized are
  \eqn\BndCond{
   ds^2 = {R^2 \over z^2} 
    \left( g_{\mu\nu} dx^\mu dx^\nu + dz^2 \right) + O(1) \qquad
    \hbox{as $z \to R$} \ .
  }
All fluctuations have to vanish as $z\rightarrow \infty$.

 A few refinements of the identification \Ident\ are worth commenting
on.  First, it is the generator of connected Green's functions which
appears on the left hand side because the supergravity action on the
right hand side is expected to follow the cluster decomposition
principle.  Second, because classical supergravity is reliable only
for a large number $N$ of coincident branes, \Ident\ can only be
expected to capture the leading large $N$ behavior.  Corrections in
$1/N$ should be obtained as loop effects when one replaces the
classical action $S$ with an effective action $\Gamma$.  This is
sensible since the dimensionless expansion parameter $\kappa^2/R^8
\sim 1/N^2$. We also note that, since $(\alpha')^2/R^4 \sim
(g^2_{YM} N)^{-1}$, 
the string theoretic $\alpha'$ corrections
to the supergravity action translate into gauge theory
corrections proportional to inverse powers of $g^2_{YM} N$. 
Finally, the fact that there is
no covariant action for type IIB supergravity does not especially
concern us: to obtain $n$-point Green's functions one is actually
considering the $n^{\rm th}$ variation of the action, which for $n>0$
can be regarded as the $(n-1)^{\rm th}$ variation of the covariant
equations of motion.

In section \TwoPt\ we will compute two-point functions 
of massless vertex operators from \Ident, compare them
with the absorption calculations in \cite{kleb,gukt,gkThree}
and find exact agreement.
However it is instructive first to examine
boundary conditions.

\subsec{Preliminary: symmetries and boundary conditions}
\subseclab\SymBC

As a preliminary it is useful to examine the appropriate boundary
conditions and how they relate to the conformal symmetry.  In this
discussion we follow the work of Brown and Henneaux \cite{hen}.  In
the consideration of geometries which are asymptotically anti-de
Sitter, one would like to have a realization of the conformal group on
the asymptotic form of the metric.  Restrictive or less restrictive
boundary conditions at small $z$ (far from the brane) corresponds, as
Brown and Henneaux point out in the case of $AdS_3$, to smaller or
larger asymptotic symmetry groups.  On an $AdS_{d+1}$ space,
  \eqn\AnyADS{
   ds^2 = G_{mn} dx^m dx^n  
     = {R^2 \over z^2} \left( -dt^2 + d\vec{x}^2 + dz^2 \right)
  }
 where now $\vec{x}$ is $d-1$ dimensional, the boundary conditions
which give the conformal group as the group of asymptotic symmetries
are 
  \eqn\AllowedSize{
   \delta G_{\mu\nu} = O(1)\ , \qquad
   \delta G_{z\mu} = O(z)\ , \qquad
   \delta G_{zz} = O(1) \ .
  }
 Our convention is to let indices $m,n$ run from $0$ to $d$ (that is,
over the full $AdS_{d+1}$ space) while $\mu,\nu$ run only from $0$ to
$d-1$ (ie excluding $z = x^d$).  Diffeomorphisms which preserve
\AllowedSize\ are specified by a vector $\zeta^m$ which for small $z$
must have the form
  \eqn\AllowedZeta{\eqalign{
   \zeta^\mu &= \xi^\mu - {z^2 \over d} \eta^{\mu\nu} 
     \partial_\nu (\xi^\kappa{}_{,\kappa}) + O(z^4)  \cr
   \zeta^z &= {z \over d} \xi^\kappa{}_{,\kappa} + O(z^3) \ .
  }}
 Here $\xi^\mu$ is allowed to depend on $t$ and $\vec{x}$ but not $r$.
\AllowedZeta\ specifies only the asymptotic form of $\zeta^m$ at large
$r$, in terms of this new vector $\xi^\mu$.

Now the condition that the variation
  \eqn\LieVar{
   \delta G_{mn} = {\cal L}_\zeta G_{mn} = 
     \zeta^k \partial_k G_{mn} + G_{kn} \partial_m \zeta^k + 
      G_{mk} \partial_n \zeta^k
  }
 be of the allowed size specified in \AllowedSize\ is equivalent to 
  \eqn\CKE{
   \xi_{\mu,\nu} + \xi_{\nu,\mu} = {2 \over d} \xi^\kappa_{,\kappa}
  }
 where now we are lowering indices on $\xi^\mu$ with the flat space
Minkowski metric $\eta_{\mu\nu}$.  Since \CKE\ is the conformal
Killing equation in $d$ dimensions ($d=4$ for the 3-brane), we see
that we indeed recover precisely the conformal group from the set of
permissible $\xi^\mu$.

The spirit of \cite{hen} is to determine the central charge of an
$AdS_3$ configuration by considering the commutator of deformations
corresponding to Virasoro generators $L_m$ and $L_{-m}$.  This method
is not applicable to higher dimensional cases because the conformal
group becomes finite, and there is apparently no way to read off a
Schwinger term from commutators of conformal transformations.
Nevertheless, the notion of central charge can be given meaning in
higher dimensional conformal field theories, either via a curved space
conformal anomaly (also called the gravitational anomaly) or as the
normalization of the two-point function of the stress energy tensor
\cite{erd}.  
We shall see in section \TwoPt\ that a calculation
reminiscent of absorption probabilities allows us to read off the two
point function of stress-energy tensors in
${\cal N} = 4$ super-Yang-Mills, and with it the
central charge.

\subsec{The two point functions}
\subseclab\TwoPt

First we consider the case of a 
minimally coupled massless
scalar propagating in the anti-de Sitter near-horizon geometry 
(one example of such a scalar is the dilaton $\phi$ \cite{kleb}).
As a further simplification
we assume for now that $\phi$ is in the $s$-wave
(that is, there is no variation over $S^5$).  Then the action becomes
  \eqn\MinAction{\eqalign{
   S &= {1 \over 2 \kappa^2} \int d^{10} x \sqrt{G} 
    \left[ \tf{1}{2} G^{MN} \partial_M \phi \partial_N \phi \right]  \cr
     &= {\pi^3 R^8 \over 4 \kappa^2} \int d^4 x \,
      \int_R^\infty {dz \over z^3} \left[ (\partial_z \phi)^2 +  
       \eta^{\mu\nu} \partial_\mu \phi \partial_\nu \phi \right] \ .
  }}
 Note that in \MinAction, as well as in all the following equations,
we take $\kappa$ to be the ten-dimensional gravitational constant.
The equations of motion resulting from the variation of $S$ are
  \eqn\MinEOM{
   \left[ z^3 \partial_z {1 \over z^3} \partial_z + 
    \eta^{\mu\nu} \partial_\mu \partial_\nu \right] \phi = 0 \ .
  }
 A complete set of normalizable solutions is
  \eqn\CompleteSet{
   \phi_k(x^\ell) = \lambda_k e^{i k \cdot x} \tilde\phi_k(z) \quad 
    \hbox{where} \quad \tilde\phi_k(z) = 
{z^2 K_2(kz)\over R^2 K_2 (k R)} \ ,
   }
$$ k^2 = \vec k^2 - \omega^2
\ .$$
We have chosen the modified Bessel function $K_2(kz)$
rather than $I_2(kz)$ because the functions $K_\nu (kz)$ fall off
exponentially for large $z$, 
while the functions $I_\nu (kz)$ grow exponentially.
In other words, the requirement of regularity at the horizon (far
down the throat) tells us which solution to keep.
A connection of this choice with the absorption calculations of
\cite{kleb} is provided by the fact that,
for time-like momenta, this is the incoming wave which
corresponds to absorption from the small $z$ region. 
$\lambda_k$ is a coupling constant, and
the normalization factor has been chosen so that $\tilde\phi_k
=1$ for $z=R$.

Let us consider a coupling 
  \eqn\WVCoup{
   S_{\rm int} = \int d^4 x \, \phi(x^\lambda) {\cal O}(x^\lambda)
  }
 in the world-volume theory.  If $\phi$ is the dilaton then according
to \cite{kleb} one would have 
${\cal O} = {1\over 4 g_{YM}^2}\tr F^2+ \ldots$.  Then
the analogue of \Ident\ is the claim that 
  \eqn\SimpIdent{
   W[\phi(x^\lambda)] = K[\phi(x^\lambda)] = S[\phi(x^\lambda,z)]
  }
 where $\phi(x^\lambda,z)$ is the unique solution of the equations of
motion with $\phi(x^\lambda,z) \to \phi(x^\lambda)$ as $z \to R$.
Note that the existence and uniqueness of $\phi$ are guaranteed
because the equation of motion is just the laplace equation on the
curved space.  (One could in fact compactify $x^\lambda$ on 
very large $T^4$ and
impose the boundary condition $\phi(x^\lambda,z) = \phi(x^\lambda)$ at
$z = R$.  Then the determination of $\phi(x^\lambda,z)$ is just the
Dirichlet problem for the laplacian on a compact manifold with
boundary).  

Analogously to the work of \cite{Polch} on the $c=1$ matrix model, we
can obtain the quadratic part of
$K[\phi(x^\lambda)]$ as a pure boundary term through
integration by parts,
  \eqn\KBound{\eqalign{
   K[\phi(x^\lambda)] &= {\pi^3 R^8 \over 4 \kappa^2} 
     \int d^4 x \, \int_R^\infty {dz \over z^3} 
     \left[ -\phi \left( z^3 \partial_z {1 \over z^3} \partial_z + 
      \eta^{\mu\nu} \partial_\mu \partial_\nu \right) \phi  + 
      z^3 \partial_z \left( \phi {1 \over z^3} \partial_z \phi \right)
     \right]  \cr
    &={1\over 2}\int d^4 k d^4 q \lambda_k \lambda_q
(2\pi)^4 \delta^4(k+q) {N^2 \over 16 \pi^2} {\cal F}
\ ,  }}
where we have expanded
$$\phi (x^\lambda) = \int d^4 k \lambda_k e^{i k\cdot x}\ .
$$ 
The ``flux factor'' ${\cal F}$ (so named because of its
resemblance to the particle number flux in a scattering calculation)
is
  \eqn\FluxDef{
   {\cal F} = 
\left[ \tilde \phi_k {1 \over z^3} \partial_z 
\tilde \phi_k \right]_R^\infty \ .
  }
 In \KBound\ we have suppressed the boundary terms in the $x^\lambda$
directions---again, one can consider these compactified on 
very large $T^4$ so
that there is no boundary.  We have also used \DThreeVars\ to simplify
the prefactor.  Finally, we have cut off the integral at $R$ as a
regulator of the small $z$ divergence.  This is in fact appropriate
since the D3-brane geometry is anti-de Sitter only for $z \gg R$.
Since there is exponential falloff in $\tilde\phi_k$ as $z\to\infty$,
only the behavior at $R$ matters.

To calculate the two-point function of ${\cal
O}$ in the world-volume theory, we differentiate $K$ twice
with respect to the coupling constants $\lambda$:\foot{The 
appearance of the logarithm 
here is analogous to the logarithmic scaling violation in the
$c=1$ matrix model.}
  \eqn\TwoPtO{\eqalign{
   \langle {\cal O}(k) {\cal O}(q) \rangle &= 
    \int d^4 x d^4 y \, e^{i k \cdot x + i q \cdot y} 
     \langle {\cal O}(x) {\cal O}(y) \rangle  \cr
    &= {\partial^2 K\over \partial \lambda_k \partial \lambda_q }=
(2\pi)^4 \delta^4(k+q) {N^2 \over 16 \pi^2} {\cal F}  \cr
    &= -(2\pi)^4 \delta^4(k+q) {N^2 \over 64 \pi^2} 
        k^4 \ln (k^2 R^2) + \hbox{(analytic in $k^2$)}
  }}
 where now the flux factor has been evaluated as
  \eqn\EvalFlux{
   {\cal F} = \left[ \tilde\phi_k
{1 \over z^3} \partial_z \tilde\phi_k 
    \right]_{z=R} =
\left[ {1 \over z^3} \partial_z  \ln (\tilde\phi_k) \right]_{z=R}
     = -\tf{1}{4} k^4 \ln (k^2 R^2) + 
      \hbox{(analytic in $k^2$)} \ .
  }
Fourier transforming back to position space, we find 
\eqn\poscorr{
\langle {\cal O} (x) {\cal O} (y) \rangle \sim {N^2\over
\vert x - y \vert ^8 }\ .
}
This is consistent with the free field result for small $g_{YM}^2 N$.
Remarkably, supergravity tells us that this formula continues to
hold as $g_{YM}^2 N\rightarrow \infty$.

Another interesting application of this analysis is to the two-point
function of the stress tensor, which with the normalization
conventions of \cite{gkThree} is
  \eqn\TwoPtFct{
   \langle T_{\alpha\beta}(x) T_{\gamma\delta}(0) \rangle
     = {c \over 48 \pi^4} X_{\alpha\beta\gamma\delta}
       \left( {1 \over x^4} \right) \ ,
  }
 where the central charge (the conformal anomaly) is $c = N^2/4$ and 
  \eqn\XFourDef{
   \vcenter{\openup1\jot
   \halign{\strut\span\TL & \span\TR\cr
     X_{\alpha\beta\gamma\delta} &= 
      2 \square^2 \eta_{\alpha\beta} \eta_{\gamma\delta} - 
       3 \square^2 (\eta_{\alpha\gamma} \eta_{\beta\delta} + 
        \eta_{\alpha\delta} \eta_{\beta\gamma}) -
      4 \partial_\alpha \partial_\beta 
        \partial_\gamma \partial_\delta  \cr
      &\quad - 2 \square 
        (\partial_\alpha \partial_\beta \eta_{\gamma\delta} +
         \partial_\alpha \partial_\gamma \eta_{\beta\delta} +
         \partial_\alpha \partial_\delta \eta_{\beta\gamma} +
         \partial_\beta \partial_\gamma \eta_{\alpha\delta} +
         \partial_\beta \partial_\delta \eta_{\alpha\gamma} +
         \partial_\gamma \partial_\delta \eta_{\alpha\beta}) \ .  \cr
   }}}
 For metric perturbations $g_{\mu\nu} = \eta_{\mu\nu} + h_{\mu\nu}$
around flat space, the coupling of $h_{\mu\nu}$ at linear order is
  \eqn\SInt{
   S_{\rm int} = \int d^4 x \, \tf{1}{2} h^{\mu\nu} T_{\mu\nu} \ .
  }
 Furthermore, at quadratic order the supergravity
action for a graviton polarized along the brane,
$h_{xy}(k)$, is exactly the minimal scalar action, 
provided the momentum $k$ is orthogonal to the $xy$ plane.
We can therefore carry over the result \TwoPtO\ to obtain
  \eqn\TwoTT{
   \langle T_{xy}(k) T_{xy}(q) \rangle 
    = - (2\pi)^4 \delta^4(k+q) {N^2 \over 64 \pi^2} 
        k^4 \ln (k^2 R^2) + 
         \hbox{(analytic in $k^2$)} \ ,
  }
 which upon Fourier transform can be compared with \TwoPtFct\ to
give $c = N^2/4$.  In view of the conformal symmetry of both the
supergravity and the gauge theory, the evaluation of this one
component is a sufficient test.

The conspiracy of overall factors to give the correct normalization of
\TwoTT\ clearly has the same origin as the successful prediction of
the minimal scalar $s$-wave absorption cross-section
\cite{kleb,gukt,gkThree}. The absorption cross-section is, up to
a constant of proportionality, the imaginary part of \TwoTT.
In \cite{kleb} the absorption cross-section was calculated in
supergravity using propagation of a scalar field in the entire
3-brane metric, including the asymptotic region far from the brane. 
Here we have, in effect, replaced communication of the throat region with
the asymptotic region by a boundary condition at one end
of the throat. The physics of this is clear: signals coming
from the asymptotic region excite the part of the throat near
$z=R$. Propagation of these excitations into the throat
can then be treated just in the anti-deSitter approximation.
Thus, to extract physics from anti-deSitter space we 
introduce a boundary at $z=R$ and take careful account of 
the boundary terms that contain the dynamical information.

It now seems clear how to proceed to three-point functions: on
the supergravity side one must expand to third order in the perturbing
fields, including in particular the three point vertices.  
At higher
orders the calculation is still simple in concept (the classical
action is minimized subject to boundary conditions), but the
complications of the ${\cal N}=8$ supergravity theory seem likely to
make the computation of, for instance, the four-point function,
rather tedious. We leave the details of such calculations for
the future. It may be very useful to compute at least the three point
functions in order to have a consistency check on the normalization of
fields.  

One extension of the present work is to consider what fields couple to
the other operators in the $N=4$ supercurrent multiplet.  The
structure of the multiplet (which includes the supercurrents, the
$SU(4)$ $R$-currents, and four spin $1/2$ and one scalar field)
suggests a coupling to the fields of gauged $N=4$ supergravity.  The
question then becomes how these fields are embedded in $N=8$
supergravity.  We leave these technical issues for the future, but
with the expectation that they are ``bound to work'' based on
supersymmetry. 

The main lesson we have extracted so far is that, for
certain operators that couple to the massless string states, the
anomalous dimensions vanish. We expect this to hold for all
vertex operators that couple to the fields of
supergravity. This may be the complete set of operators
that are protected by supersymmetry. As we will
see in the next section, other operators acquire anomalous dimensions
that grow for large `t Hooft coupling.

\newsec{Massive string states and anomalous dimensions}
\seclab\MassiveString

Before we proceed to the massive string states, 
a useful preliminary is to discuss the higher partial waves of a
minimally coupled massless scalar.
The action in five
dimensions (with Lorentzian signature), the equations of motion, and
the solutions are
  \eqn\PartAction{
   S = {\pi^3 R^8 \over 4 \kappa^2} \int d^4 x \int_R^\infty
        {dz \over z^3} \left[ (\partial_z \phi)^2 +
         (\partial_\mu \phi)^2 + {\ell (\ell + 4) \over z^2} \phi^2
        \right]
  }
  \eqn\PartEOM{
   \left[ z^3 \partial_z {1 \over z^3} \partial_z +
    \eta^{\mu\nu} \partial_\mu \partial_\nu -
    {\ell (\ell + 4) \over z^2} \right] \phi = 0
  }
  \eqn\PartSols{
   \phi_k(x^\ell) = e^{i k \cdot x} \tilde\phi_k(z) \quad
    \hbox{where} \quad
   \tilde\phi_k(z) = {z^2 K_{\ell+2}(kz) \over
    R^2 K_{\ell+2}(kR)} \ .
  }
We have chosen the normalization such that
$\tilde\phi_k(z) =1$ at $z=R$.
The flux factor is evaluated by expanding
  \eqn\PhiBehave{
   K_{\ell+2}(kz) = 2^{\ell+1}\Gamma(\ell+2)
(k z)^{-(\ell+2)} \left (1 + \ldots +
     {(-1)^\ell \over 2^{2\ell+3} (\ell+1)! (\ell+2)!}
     (kz)^{2(\ell+ 2)} \ln kz + \ldots \right )\ ,
  }
where in parenthesis we exhibit the leading non-analytic term.
We find
  \eqn\GotFlux{
   {\cal F} =
\left[ {1 \over z^3} \partial_z  \ln (\tilde\phi_k) \right]_{z=R}=
 {(-1)^\ell \over 2^{2 \ell + 2}
    [(\ell+1)!]^2 } k^{4+ 2l} R^{2\ell} \ln kR \ .
  }
 As before, we have neglected terms containing analytic powers of
$k$ and focused on the leading nonanalytic term.
This formula indicates that the operator that couples to $\ell$-th
partial wave has dimension $4+\ell$.
In \cite{kleb} it was shown that such operators with the $SO(6)$
quantum numbers of the $\ell$-th partial wave have the form
\eqn\higher{
\int d^4 x e^{ik\cdot x} \tr \left [ \left (X^{(i_i}\ldots X^{i_\ell )}
+ \ldots \right )
F_{\mu\nu} F^{\mu\nu} (x) \right ]\ , }
where in parenthesis we have a traceless symmetric tensor of $SO(6)$.
Thus, supergravity predicts that
their non-perturbative dimensions equal their bare dimensions.

Now let us consider massive string states.  
Our goal is to use supergravity to calculate the anomalous dimensions of
the gauge theory operators that couple to them.
To simplify
the discussion, let us focus on excited string states which are
spacetime scalars of mass $m$.  The propagation equation for such a
field in the background of the 3-brane geometry is
\eqn\Coulone{
\left [{d^2\over d
r^2} + {5\over r} {d\over dr} - k^2 \left (1 + {R^4\over r^4}\right )
-m^2 \left (1 + {R^4\over r^4}\right )^{1/2} \right ]
\tilde \phi_k =0\ . } 
For the state at excitation level $n$,
$$ m^2 = {4 n\over
\alpha'} \ . $$ 
In the throat region, $z\gg R$, \Coulone\ simplifies to 
\eqn\Coultwo{ \left [{d^2\over d
z^2} - {3\over z} {d\over dz} -k^2 
-{m^2 R^2\over z^2} \right ]\tilde \phi_k =0\ . } 
Note that a massive particle with small energy $\omega \ll m$, which would
be far off shell in the asymptotic region $z\ll R$, can nevertheless
propagate in the throat region (i.e. it is described by an oscillatory
wave function).
 
Equation \Coultwo\
is identical to the equation encountered in the analysis of higher
partial waves, except
the effective angular momentum is not in general an integer:
in the centrifugal barrier term $\ell (\ell+4)$ is replaced by $(m R)^2$.
Analysis of the choice of wave function goes through as before,
with $\ell+2$ replaced by $\nu$, where
\eqn\anom{\nu = \sqrt{ 4 + (m R)^2 } \ .}
In other words, the wave function falling off exponentially for
large $z$ and normalized to $1$ at $z=R$ is\foot{
These solutions are reminiscent of the loop correlators
calculated for $c\leq 1$ matrix models in \cite{mss}.}
\eqn\massivewave{
   \tilde\phi_k(z) = {z^2 K_\nu(kz) \over
    R^2 K_\nu (k R)} \ .}
Now we recall that
$$ K_\nu = {\pi\over 2\sin (\pi\nu)} \left (I_{-\nu} - I_\nu \right )\ ,
$$
$$ I_\nu (z) = \left ({z\over 2}\right )^\nu
\sum_{k=0}^\infty {(z/2)^{2k} \over k! \Gamma (k+\nu+ 1)}\ .
$$
Thus, 
\eqn\genexpand{ K_\nu (kz) = 2^{\nu -1}\Gamma (\nu)
(k z)^{-\nu} \left (1+ \ldots -  
\left ({z k\over 2 }\right )^{2\nu } 
{\Gamma (1-\nu)\over \Gamma (1+\nu )}+\ldots \right )\ ,
} 
where in parenthesis we have exhibited the leading non-analytic term.
Calculating the flux factor as before, 
we find that the leading non-analytic term of the
2-point function is
  \eqn\OOFct{
   \langle {\cal O}(k) {\cal O}(q) \rangle =-(2\pi)^4 \delta^4(k+q)
{N^2 \over 8\pi^2} {\Gamma (1-\nu)\over \Gamma (\nu )}
    \left ({kR\over 2}\right )^{2\nu} R^{-4}\ .
  }
This implies that the dimension of the corresponding SYM
operator is equal to 
\eqn\opdim{
\Delta= 2+\nu= 2+\sqrt{ 4 + (m R)^2 }\ .
}

Now, let us note that
$$ R^4 = 2 N g_{YM}^2 (\alpha')^2\ ,
$$
which implies
$$ (m R)^2 = 4 n g_{YM} \sqrt {2 N} 
\ .
$$
Using \anom\ we find that the spectrum of dimensions
for operators that couple to massive string states is,
for large $g_{YM} \sqrt N$,
\eqn\anomform{ 
h_n \approx 2\left (n g_{YM} \sqrt {2 N} \right )^{1/2}
\ .}
Equation \anomform\ is a new 
non-trivial prediction of the string theoretic
approach to large $N$ gauge theory.\foot{We do not expect this equation
to be valid for arbitrarily large $n$, because application of linearized
local effective actions to arbitrarily excited string states 
is questionable.
However, we should be able to trust our approach for moderately excited
states.}

We conclude that, for large `t Hooft coupling,
the anomalous dimensions of the
vertex operators corresponding to massive string
states grow without bound.
By contrast, the vertex operators that couple to the massless string
states do not acquire any anomalous dimensions. This has been checked
explicitly for gravitons, dilatons and RR scalars \cite{kleb,gukt,gkThree},
and we believe this to be a general statement.\foot{A more general
set of such massless fields is contained in the supermultiplet of 
AdS gauge fields, whose boundary couplings were recently
studied in \cite{ff}.} Thus, there are several $SO(6)$ towers of
operators that do not acquire anomalous dimensions, such as
the dilaton tower \higher. The absense of
the anomalous dimensions
is probably due to the fact that they are protected by SUSY. The rest
of the operators are not protected and can receive arbitrarily large
anomalous dimensions. While we cannot yet write down the
explicit form of these operators in the gauge theory, 
it seems likely that they
are the conventional local operators, such as 
\gaugeop.  Indeed,
the coupling of a highly excited string state to the world volume may
be guessed on physical grounds.  Since a string in a D3-brane is a
path of electric flux, it is natural to assume that a string state
couples to a Wilson loop ${\cal O} = \exp \left( i \oint_\gamma A
\right)$. Expansion of a small loop in powers of $F$ yields
the local polynomial operators.

There is one potential problem with our treatment of massive
states. The minimal linear equation \Coulone\ where higher
derivative terms are absent may be true only for a particular field
definition (otherwise corrections in positive
powers of $\alpha' \nabla^2$ will be present in the equation).
Therefore, it is possible that there are energy dependent leg
factors relating the operators $\cal O$ in 
\OOFct\ and the gauge theory operators of the form \gaugeop.
We hope that these leg factors do not change our conclusion about the
anomalous dimensions. However, to completely settle this issue we
need to either find an exact sigma-model which incorporates all
$\alpha'$ corrections or to calculate the 3-point functions
of massive vertex operators.

\newsec{Conclusions}

There are many unanswered questions that we have left for the future.
So far, we have considered the limit of large `t Hooft coupling,
since we used the one-loop sigma-model calculations for
all operators involved. If this coupling is not large, then we have
to treat the world sheet theory as an exact conformal field theory
(we stress, once again, that the string loop corrections are
$\sim 1/N^2$ and, therefore, vanish in the large $N$ limit).
This conformal field theory is a sigma-model on a hyperboloid.
It is plausible that, in addition to the global $O(2,4)$ symmetry,
this theory possesses the $O(2,4)$ Kac-Moody algebra.
If this is the case, then the sigma model is tractable
with standard methods of conformal field theory.

Throughout this work we detected many formal similarities of our
approach with that used in $c\leq 1$ matrix models. These models
may be viewed as early examples of gauge theory -- non-critical string
correspondence, with the large $N$ matrix models playing the role
of gauge theories. Clearly, a deeper understanding of
the connection between the present work and the $c\leq 1$ matrix models
is desirable.

\bigbreak\bigskip\bigskip\centerline{{\bf Acknowledgements}}\nobreak
\bigskip

We would like to thank C.~Callan and the participants of the
``Dualities in String Theory'' program at ITP, Santa Barbara, for useful
discussions. I.R.K. is grateful to ITP for
hospitality during the completion of this paper.
This research was supported in part by the National Science 
Foundation under Grants No. PHY94-07194 and
PHY96-00258, by the Department
of Energy under Grant No.  
DE-FG02-91ER40671, 
and by the James S.~McDonnell Foundation under Grant No. 91-48.  
S.S.G. also thanks the Hertz Foundation for its support.

\bigskip
\bigskip

\centerline{\bf References}
\bigskip

\bibliography{ads}   
\bibliographystyle{ssg}

\bye